\documentclass[11pt]{article}
\usepackage{moriond,epsfig}

\newcommand{\fbi}{\ensuremath{\,\mathrm{fb}^{-1}}}
\newcommand{\pbi}{\ensuremath{\,\mathrm{pb}^{-1}}}

\newcommand{\Fig}{Fig.}

\begin{document}
\vspace*{4cm}
\title{Electroweak Measurements with the ATLAS and CMS Experiments}

\author{ J. Kretzschmar on behalf of the ATLAS and CMS
Collaborations}

\address{Department of Physics, Oxford St,
Liverpool, L69 7ZE, UK}

\maketitle\abstracts{ Highlights of ATLAS and CMS measurements
  involving the production of heavy electroweak gauge bosons, $W$ and
  $Z$, at the LHC are presented. Cross sections of single $W$ and $Z$
  bosons are studied with very high precision and differential in
  various kinematic variables. The rapidity differential measurements
  are shown to have a so far unique impact on our knowledge of proton
  structure with regards to the strange quark density. The production
  in association with one or more light or heavy flavour jets is
  studied. Furthermore measurements of $\tau$ final states, $W$
  polarisation and the weak mixing angle $\sin^2\theta_W$ are
  presented. Various di-boson measurements are presented and
  measurements are in general found to be well described by the
  Standard Model predictions. These measurements test the non-Abelian
  gauge structure and limits on anomalous triple gauge couplings are
  derived, which are of impact comparable to the corresponding LEP and
  Tevatron results.}

\section{Introduction}

The first two years of operation of the Large Hadron Collider (LHC),
2010 and 2011, have enabled the ATLAS~\cite{Aad:2008zzm} and
CMS~\cite{Chatrchyan:2008aa} collaboration to study a wealth of
processes involving the production of the heavy electroweak bosons,
$W$ and $Z$. These cover a wide range in cross sections, starting from
the inclusive production of single $W$ and $Z$ bosons and ending with
the rather rare production of di-boson final states like $ZZ$. The
results are of exceptional quality thanks to well understood detectors
and quickly increasing integrated luminosity with about $40\pbi$
collected in 2010 and already $5\fbi$ in 2011.

The study of electroweak processes is indispensable to understand
the background to new physics signals. But accurate cross section
measurements are interesting in their own right.

First, single $W$ and $Z$ boson production is a very sensitive probe
of Quantum Chromodynamics (QCD). Inclusive production as well as
production in association with heavy flavours is sensitive to the
proton structure as described by Parton Distribution Functions (PDFs).
Production in association with many jets or at high boson transverse
momentum challenges our understanding of perturbative QCD calculations
in extreme configurations. 

Second, the single or double production of the electroweak
bosons is used to study their properties in details, for example their
couplings to fermions or other gauge bosons. The self-couplings of
the bosons are predicted by the Standard Model and non-zero
anomalous triple gauge couplings would open a window to new Physics.

Due to the wealth of results only some of these can be highlighted here.

\section{Production of Single $W$ and $Z$ Bosons}

\begin{figure}
  \begin{center}
    \psfig{figure=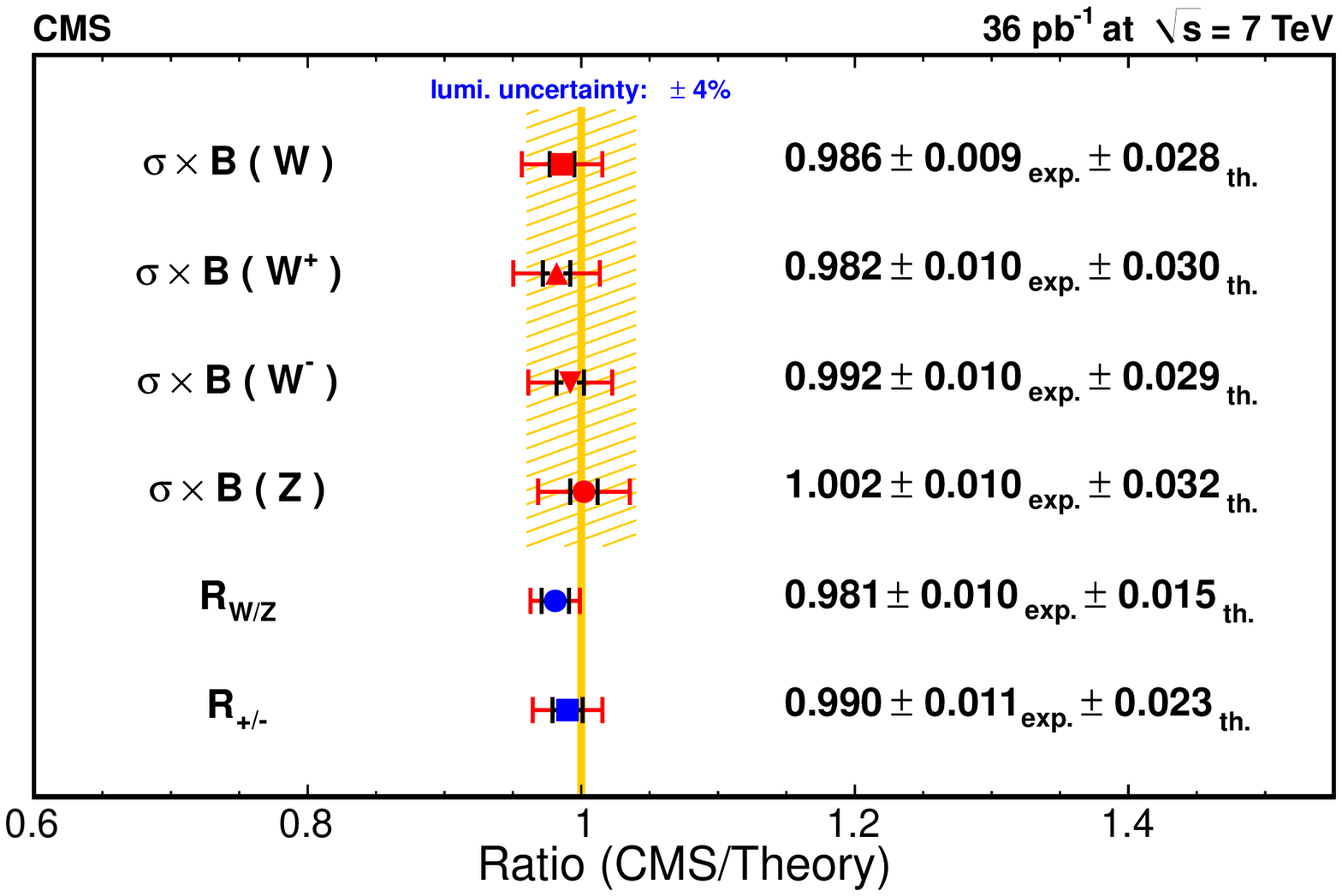,width=0.485\linewidth}%
    \psfig{figure=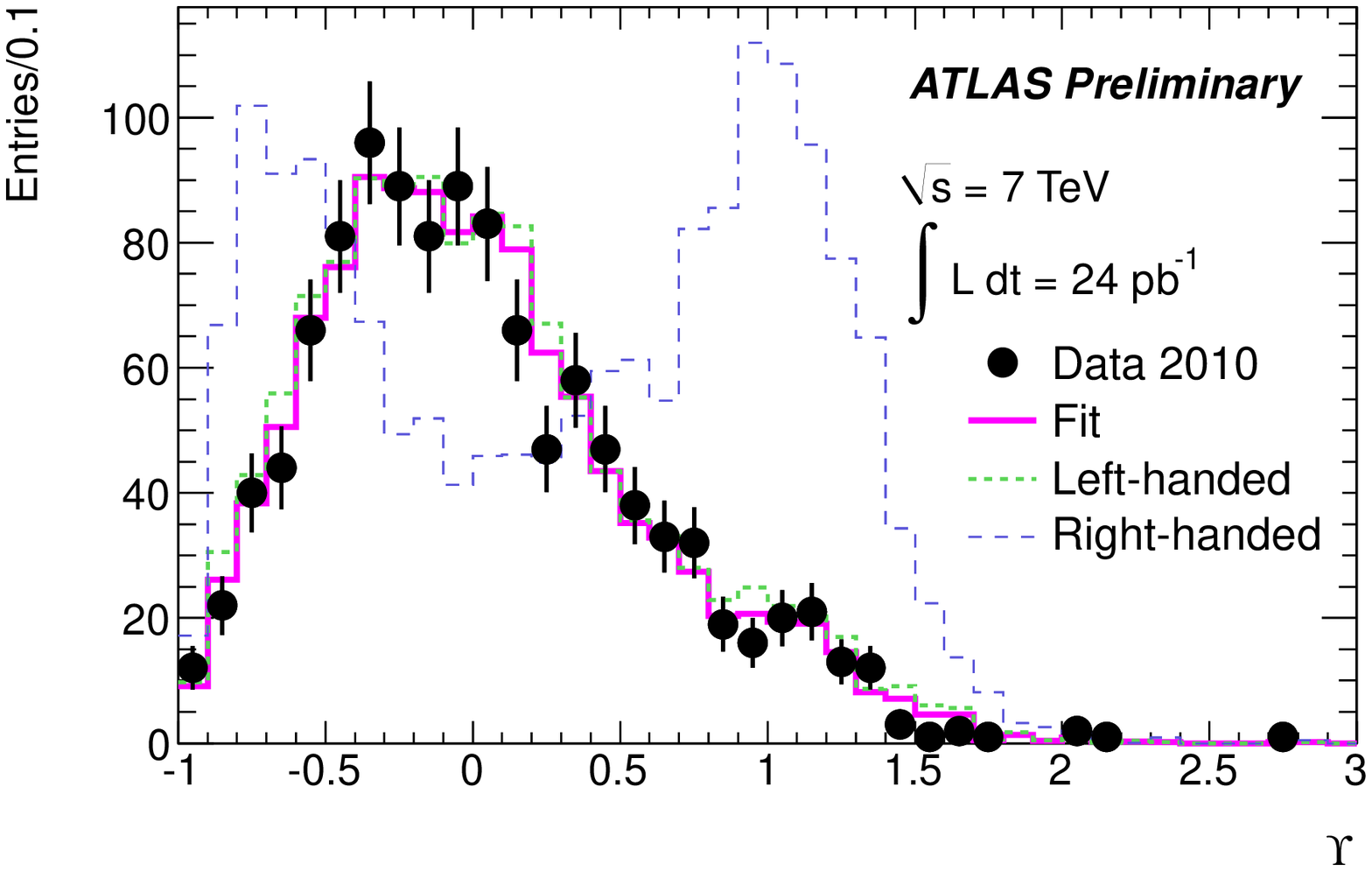,width=0.515\linewidth}
  \end{center}
  \caption{Left: Example of total integrated $W^\pm$ and $Z$ cross
    section measurements and their ratios, compared to theory
    prediction at NNLO. Very good overall agreement can be observed~\protect\cite{CMS:2011aa}.
    Right: Energy sharing variable $\Upsilon$ in
    hadronic one prong $\tau$ decays from the process $W\to\tau\nu$.
    The data is compared to templates corresponding to right or
    left-handed $\tau$ polarisation, where the expected left-handed
    polarisation is clearly favoured~\protect\cite{Aad:2012cu}.
    \label{fig:wzint_taupol}}
\end{figure}

The inclusive production and leptonic decays of the heavy electroweak
bosons, $W\to \ell\nu$ and $Z \to \ell\ell$ in the electron and muon
channels, $\ell = e, \mu$, are among the standard candles at hadron
colliders. Already with the 2010 data an experimental precision of
$1-2\%$ was reached~\cite{Aad:2011dm,CMS:2011aa}. These
measurements constitute a precision test of QCD at NNLO and are
sensitive to the proton PDFs. The results for integrated cross
sections and their ratios are in general in good agreement with the
theory predictions, as shown for example in
\Fig~\ref{fig:wzint_taupol}, left.

The $W$ and $Z$ cross sections have also been measured in $\tau$
decays~\cite{Aad:2011kt,Ztau2011,Chatrchyan:2011nv,Aad:2011fu,CMS-PAS-EWK-11-019},
where the cross section precision is limited to typically $10-20\%$. An interesting recent
application is the first measurement of the $\tau$ polarisation in $W$
decays at a hadron collider\cite{Aad:2012cu}. Hadronic one-prong
decays of the $\tau$ boson are analysed concerning the energy sharing
between neutral and charged decay products, $\Upsilon = (E_T^{\pi-} -
E_T^{\pi0}) / (E_T^{\pi-} + E_T^{\pi0})$, which has a strong
dependence on the polarisation, see \Fig~\ref{fig:wzint_taupol},
right. The measured polarisation of $P_\tau = -1.06 \pm
0.04_\mathrm{stat} \, ^{+0.05}_{-0.07} \,_\mathrm{syst}$ is compatible
with the SM expectation of $P_\tau=-1$ and proves that this technique
may be applied to determine spin properties of new particles decaying
to $\tau$ final states.

\begin{figure}
  \begin{center}
    \psfig{figure=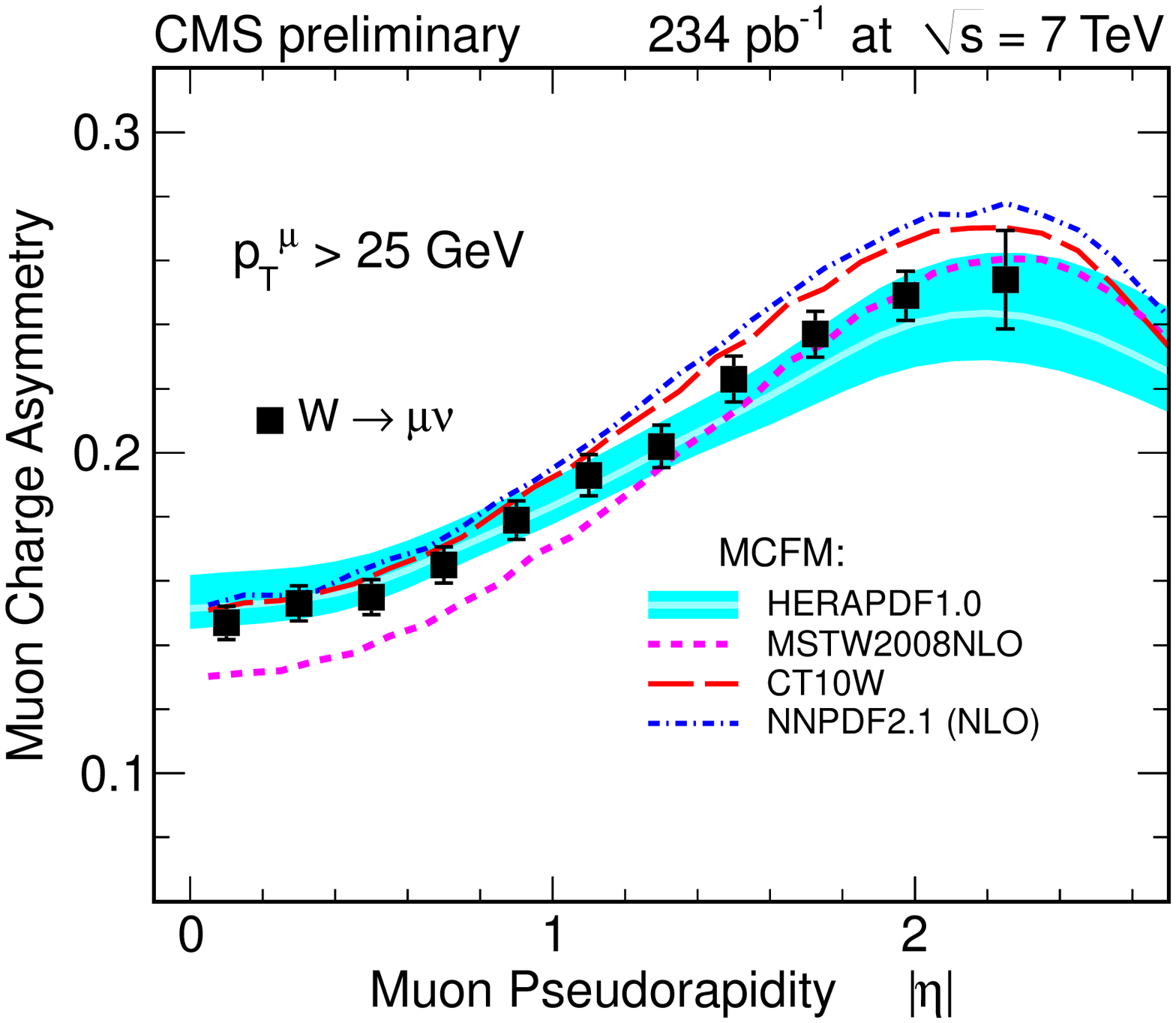,width=0.5\linewidth}%
    \hspace{0.05\linewidth}%
    \psfig{figure=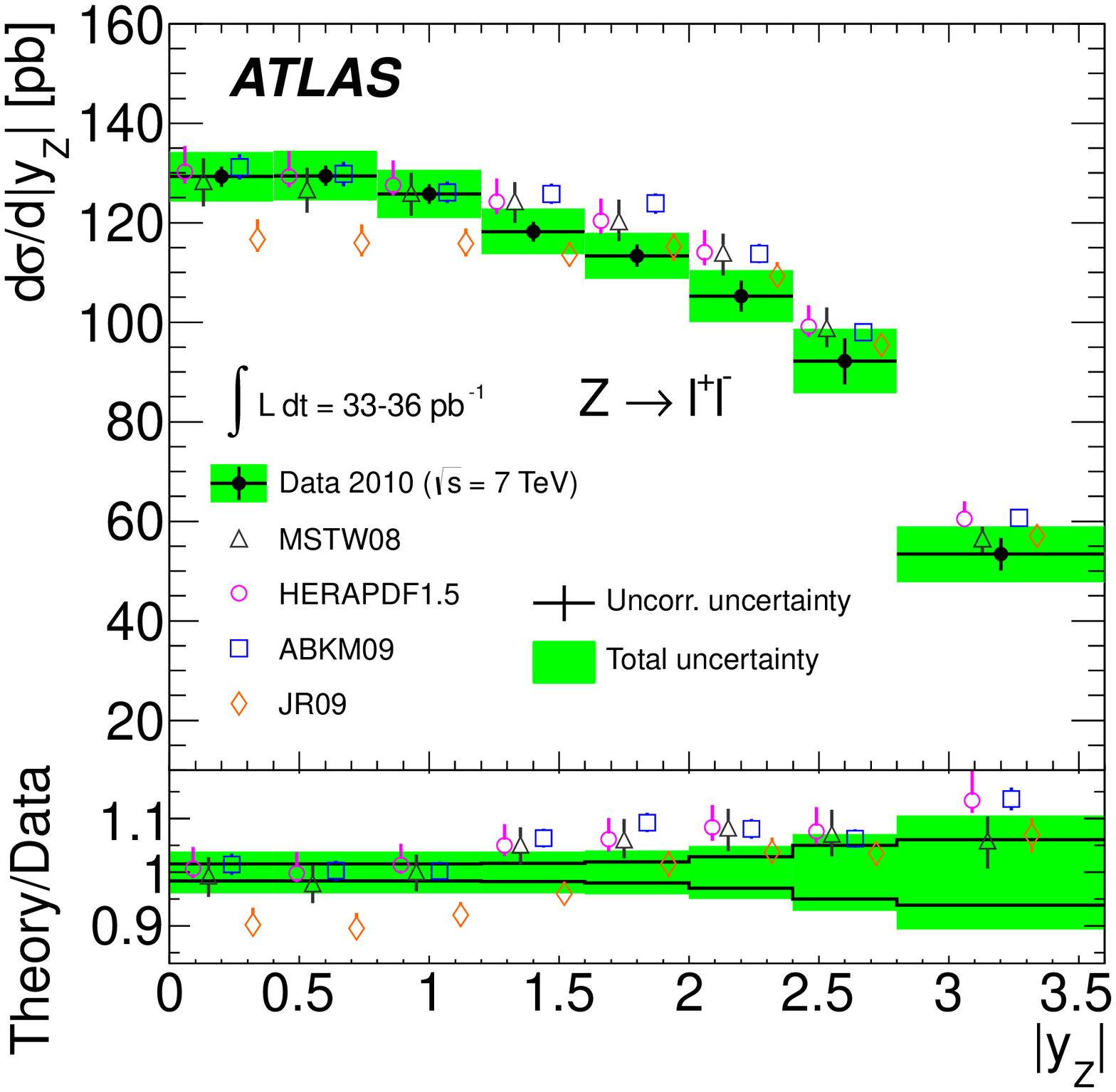,width=0.45\linewidth}
  \end{center}
  \caption{Left: $W\to\mu\nu$ charge asymmetry as function on muon
    pseudo-rapidity compared to the predictions with four different PDF
    sets~\protect\cite{CMS-PAS-EWK-11-005}. Right: Boson rapidity differential cross section for the
    process $Z\to\ell\ell$ compared to the predictions with four PDF
    sets~\protect\cite{Aad:2011dm}. Some PDF sets are clearly favoured over others by these
    measurements. \label{fig:wzdiff}}
\end{figure}

Constraints on the PDFs from $W$ and $Z$ production is maximised using
differential cross section information. The boson rapidity $y$ is
directly linked to parton momentum fractions as $x_{1,2} = M_{W,
  Z}/\sqrt{s} \cdot e^{\pm y}$. As for the $W$ the boson rapidity
cannot be reconstructed, the charged lepton pseudo-rapidity
$\eta_\ell$ used instead to yield correlated information. The CMS
collaboration has measured the $W$ lepton charge asymmetry, $A(\eta) =
(\mathrm{d}\sigma^+(\eta) -
\mathrm{d}\sigma^-(\eta))/(\mathrm{d}\sigma^+(\eta) +
\mathrm{d}\sigma^-(\eta))$~\cite{Chatrchyan:2011jz,CMS-PAS-EWK-11-005}
as well as the normalised $Z$ rapidity distribution $1/\sigma \cdot
\mathrm{d}\sigma/\mathrm{d}y$~\cite{Chatrchyan:2011wt}. The ATLAS
collaboration has instead measured absolute differential cross
sections for $Z$, $W^+$ and $W^-$ with the full uncertainty
correlation information~\cite{Aad:2011dm}. A few examples of these
measurements and theory comparisons at NLO or NNLO with various PDF
sets are given in
\Fig~\ref{fig:wzdiff}. A broad agreement between measurement and
predictions can be seen, however there are also significant
differences, which indicate a sensitivity of the measurements to the
PDFs. Meanwhile CMS has made public new results on differential
$Z/\gamma^*\to\mu\mu$ cross sections and $W\to e\nu$ asymmetries~\cite{CMS-PAS-EWK-11-007,Chatrchyan:2012xt}.

\begin{figure}
  \begin{center}
    \psfig{figure=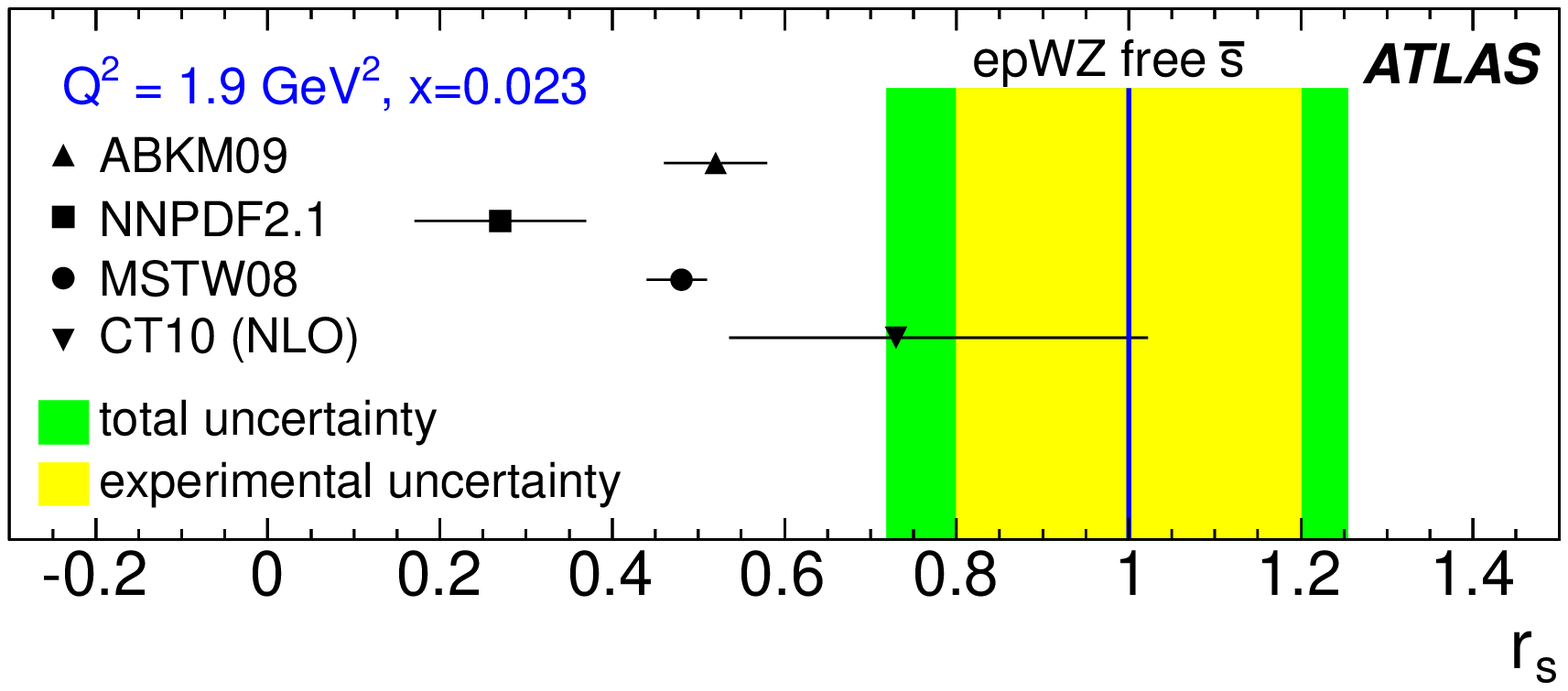,width=0.55\linewidth}%
    \psfig{figure=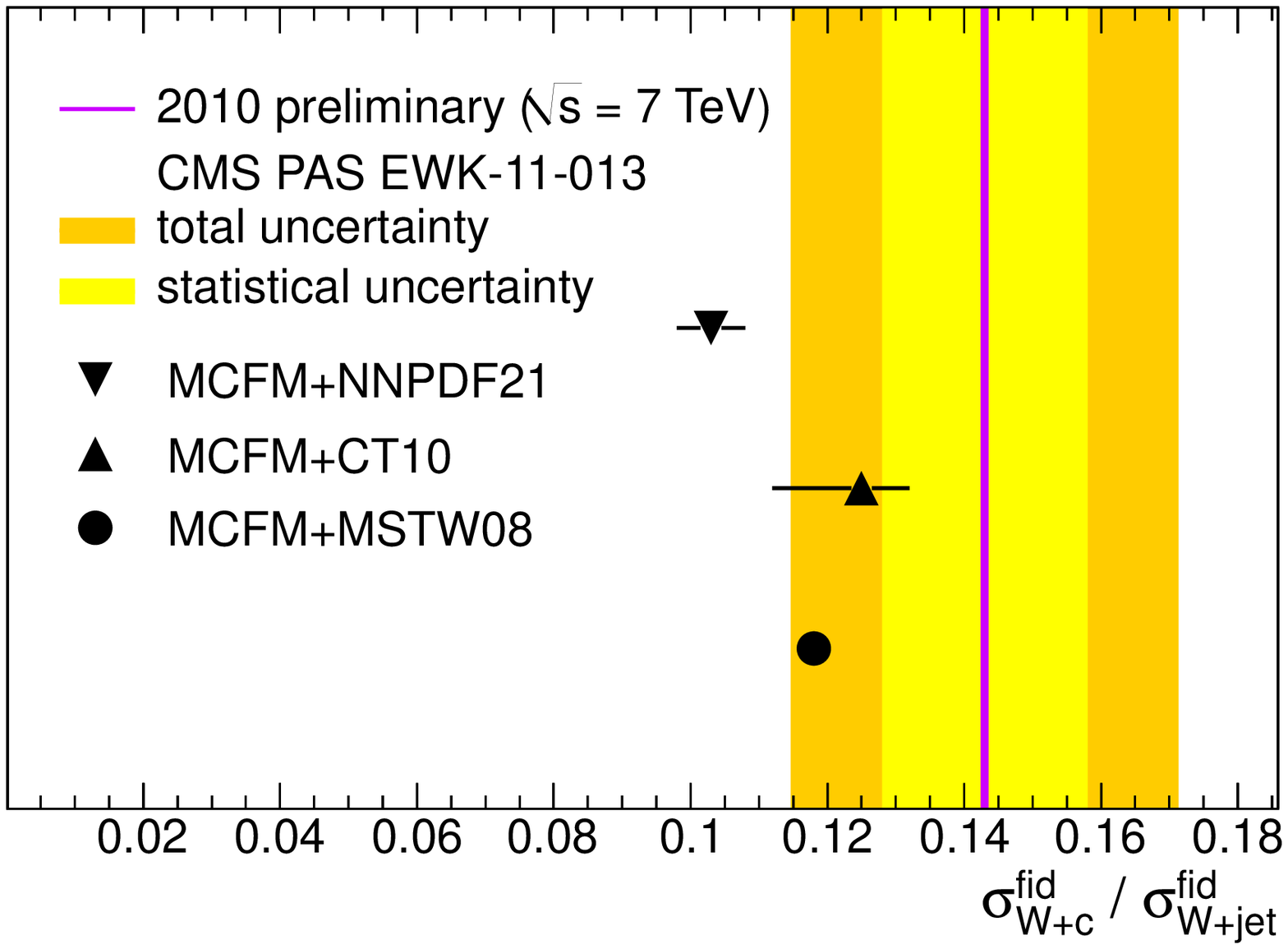,width=0.45\linewidth}
    \caption{Left: Determination of the ratio of strange to the down
      quark sea, $r_s$, compared to the predictions with four
      different PDF sets~\protect\cite{Aad:2012sb}. Right: Measurement of the ratio of of
      $W$+charm jet to $W$ plus any jet compared to the predictions with three
      different PDF sets. In both analyses a significant tendency for
      larger than expected strange quark content of the proton can be observed. \label{fig:rs_wc}}
  \end{center}
\end{figure}

ATLAS has recently performed a QCD fit at NNLO to their 2010
differential $W$ and $Z$ cross section to demonstrate the 
impact of the new data on the PDF fits~\cite{Aad:2012sb}. The
fit also uses $ep$ Deep Inelastic Scattering cross section data from
the HERA experiments in a setup similar to the one used in the HERAPDF
fit~\cite{Aaron:2009aa}. It is found, that especially the ATLAS $Z$
data has a sensitivity to the strange content of the
proton. The strange density is quantified via its ratio to the down quark
sea, $r_s=0.5 \cdot (xs(x)+x\bar{s}(x))/x\bar{d}(x)$. It is
often assumed, that the strange quark density is suppressed at low
scales $Q^2$, $r_s<1$. With a conventional value of $r_s=0.5$ at
$Q^2=1.9\,\mathrm{GeV}^2$ there is a significant tension observed and
the $\chi^2/N_\mathrm{DF} = 44.5/30$ for the ATLAS data is not very good.
Leaving the strange sea free, improves the $\chi^2$ by more than $10$
units and gives a result consistent with no strange suppression,
$r_s = 1.00 \pm 0.20_\mathrm{exp} \,^{+0.16}_{-0.20} \,_\mathrm{sys}$
at $Q^2=1.9\,\mathrm{GeV}^2$ and $x=0.023$. While the $W^\pm$ data have little
direct sensitivity to the strange, they improve the impact of the
ATLAS data, as the fit has to find a solution compatible with the
absolute normalisations of all the data points.
Figure~\ref{fig:rs_wc}, left, compares this determination to the
values of $r_s$ extracted from other global PDF fits. Only the CT10
PDF set predicts a similarly large strange content, while for other
sets a $\sim 2\sigma$ tension is observed.

A complementary approach to determine the strange content of the
proton is available through measuring $W$ production in association
with charm jets. This gives a direct access to strange quark
content, as the final state is dominated by contributions from the
Cabibbo-favoured processes $\bar{s}g \to W^+\bar{c}$ and $sg \to
W^-c$. CMS has performed first ratio measurements of these final
states using a $W+\mathrm{jet}$ selection and additional secondary
vertex tagging~\cite{CMS-PAS-EWK-11-013}. For example the ratio of
$c$-tagged over all jets was determined to be
$R_c=\sigma(Wc)/\sigma(W+\mathrm{jet}) = 0.143 \pm
0.015\,_\mathrm{stat} \pm 0.024\,_\mathrm{sys}$. As can be seen in
\Fig~\ref{fig:rs_wc} right, this measurement is in agreement with various
global PDF sets, although there is an indication of a larger than
expected strange density at the $\sim 1\sigma$ level.

\begin{figure}
  \begin{center}
    \psfig{figure=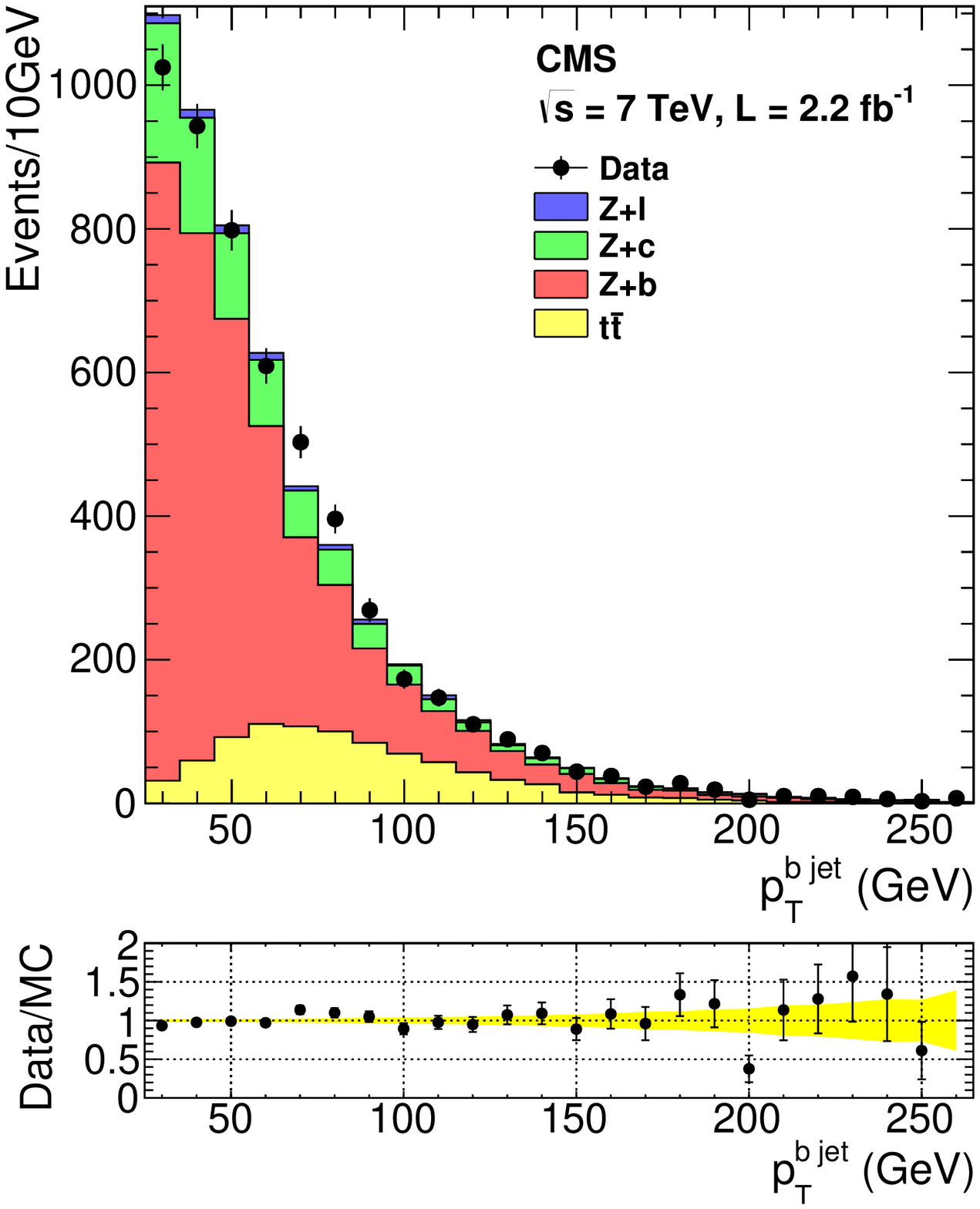,width=0.44\linewidth}
    \psfig{figure=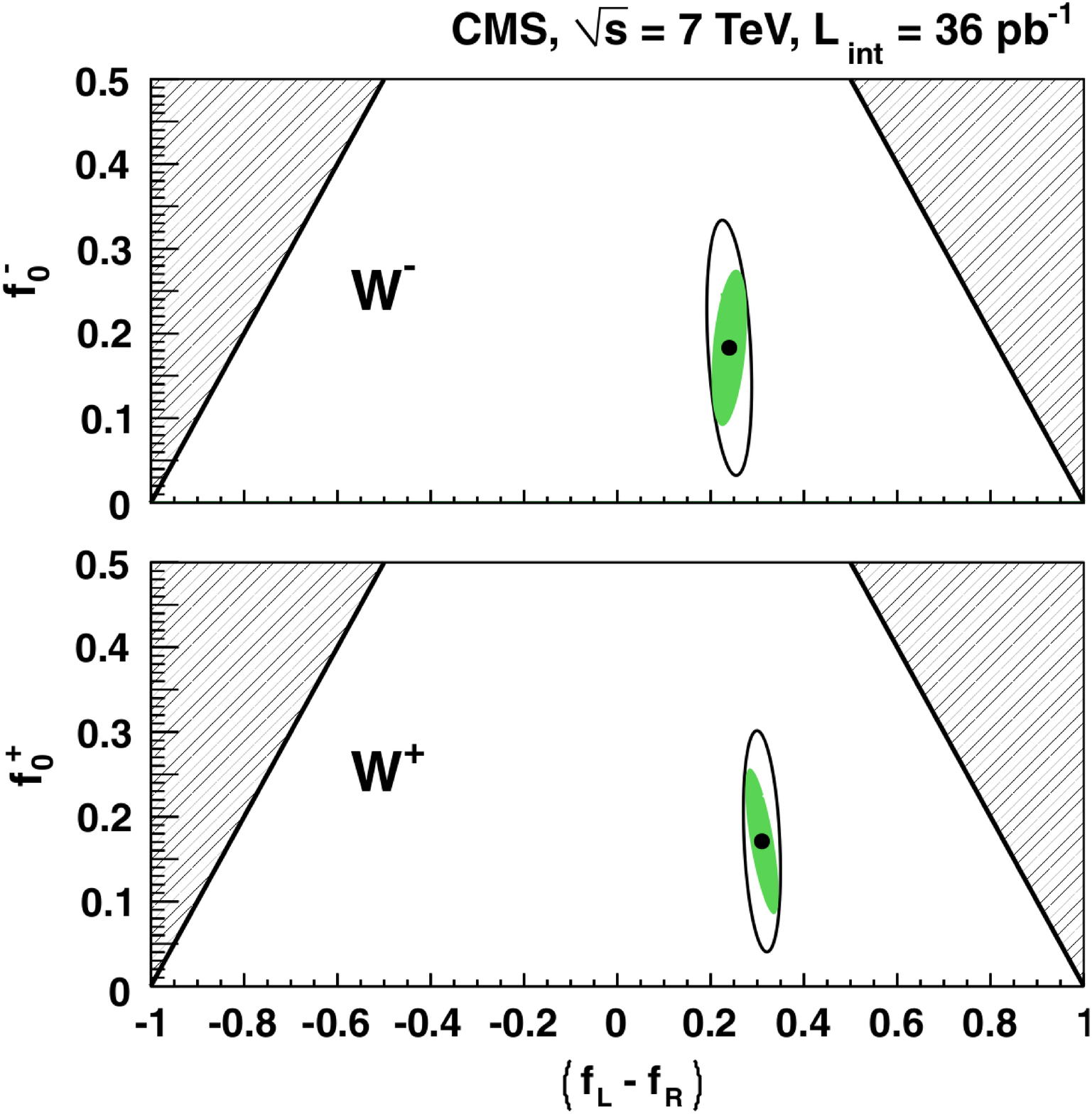,width=0.54\linewidth}
  \end{center}
  \caption{Left: Measurement of the transverse momentum spectrum of
    $b$ jets produced in association with a $Z$ boson
    compared to the MadGraph+Pythia expectation. A reasonable, but not
    perfect agreement can be observed~\protect\cite{CMS-PAS-SMP-12-003}. Right: Measurement of the $W^-$
    and $W^+$ boson polarisation at high transverse boson momentum
    $p_{T,W}>50\,$ GeV. The expectation of predominantly left-handed
    and non-zero longitudinal $W$ production are confirmed.
    \label{fig:zbb_wpol}}
\end{figure}

The production of $W$ and $Z$ in association of $b$-jets is a further
interesting test of QCD, determines a background to many new physics
searches, and is potentially sensitive to the $b$ density in the proton. ATLAS
has measured the $W+b$ process~\cite{Aad:2011kp} and both collaborations have
measured $Z+b$ processes~\cite{Aad:2011jn,CMS-PAS-SMP-12-003}. One
example of a recent CMS result on the $Z+b$ transverse momentum
spectrum is presented in \Fig~\ref{fig:zbb_wpol}, left. It can be seen,
that the shape of the distribution is in fair agreement with the
MadGraph+Pythia generator, while the integrated measured cross section
is above the MCFM NLO prediction by more than $1\sigma$.

The production of $W$ and $Z$ bosons in association with up to $4$
jets can be compared to
various tree-level matrix element generators like Alpgen, MadGraph or
Sherpa, or the NLO QCD calculation provided by BlackHat+Sherpa. Both
ATLAS and CMS have published many different cross sections and
ratios~\cite{Aad:2011qv,Aad:2012en,Aad:2011xn,Chatrchyan:2011ne}. In
general a very good agreement of all the data with the MC or
the NLO QCD predictions is found.

Massive spin 1 vector bosons like the $W$ can be produced in three
polarisation states: left- or right-handed or in longitudinal state.
The corresponding fractions are denoted as $f_L$, $f_R$, and $f_0$ and
can be measured by analysing the lepton transverse and angular
momenta. Using only transverse measurements, a quantity like $L_P =
\vec{p}_T^{\,\ell} \cdot \vec{p}_T^{\,W}/|\vec{p}_T^{\,W}|^2$ can be
defined to approximate the polarisation angle. ATLAS and CMS have
measured the fractions $f_L-f_R$ and $f_0$ for significant $W$ boson
transverse momentum $p_{T,W}$~\cite{ATLAS:2012au,Chatrchyan:2011ig}.
The CMS results for $W^+$ and $W^-$ and $p_{T,W}>50\,$GeV are shown in
\Fig~\ref{fig:zbb_wpol}, right. The predominantly left-handed $W$
production and non-zero longitudinal component as predicted by NLO QCD
is confirmed.

Finally, the large amount of $Z/\gamma^*$ bosons produced at the LHC
can be used to determine the weak mixing angle $\sin^2\theta_W$, one
of the fundamental parameters of the Standard Model. The world average
of this quantity has reached a $\sim 0.1\%$ relative
uncertainty, but there is a significant tension between the results
entering the combination. CMS has measured the effective mixing angle
from the forward-backward asymmetry in the $q\bar{q} \to Z/\gamma^*
\to \mu^-\mu^+$ process~\cite{Chatrchyan:2011ya}. Although the
incoming quark/anti-quark direction is unknown on an event-by-event
basis, it can be disentangled on a statistical basis for non-zero
boost, i.e. non-zero boson rapidity $y$. The analysis uses a three
dimensional fit in decay angle $\cos\theta^*$, mass $m$ and rapidity
$y$. The result reaches a $\sim 1\%$ precision:
$\sin^2\theta_\mathrm{eff}=0.2287 \pm 0.0020_\mathrm{stat} \pm
0.0025_\mathrm{syst}$.

\section{Di-Boson Production}

The main processes contributing to final states with multiple gauge
bosons, $W^\pm$, $Z$ or $\gamma$, are in general $t$-channel quark
exchange diagrams. In the case of final state photons, QED Final State
Radiation (FSR) is also significant. Finally, due to the non-Abelian
gauge structure of the Standard Model, $SU(2)_L \times U(1)_Y$, there
are also $s$-channel diagrams. These triple gauge couplings (TGC) are
however only present for vertices involving a $W$ boson and are zero
for the ``neutral'' vertices involving only $Z$ or $\gamma$ bosons. It
is possible to enhance the TGC contribution, typically by requiring
very high transverse momentum objects. Limits on TGCs different from
the SM prediction, so called anomalous TGCs (aTGCs), can be set. The
published ATLAS and CMS limits on aTGCs are typically derived from
data sets of of $\sim 1\fbi$. The limits are on a similar level as
obtained from Tevatron and LEP analyses, where the details depend on
the specific coupling, channel and model assumptions in the
extraction. Finally, the cross sections for di-boson production can be
compared to the NLO predictions. This also serves to understand
important irreducible background to many searches for the SM Higgs
boson.

Experimentally the measurement of di-boson production is more
challenging than for single bosons due to the smaller cross sections
and diverse processes acting as source of background. Concentrating on
the leptonic decay channels and thanks to the excellent lepton
identification capabilities, the signal is usually quite well
separated from the background.

\begin{figure}
  \begin{center}
    \psfig{figure=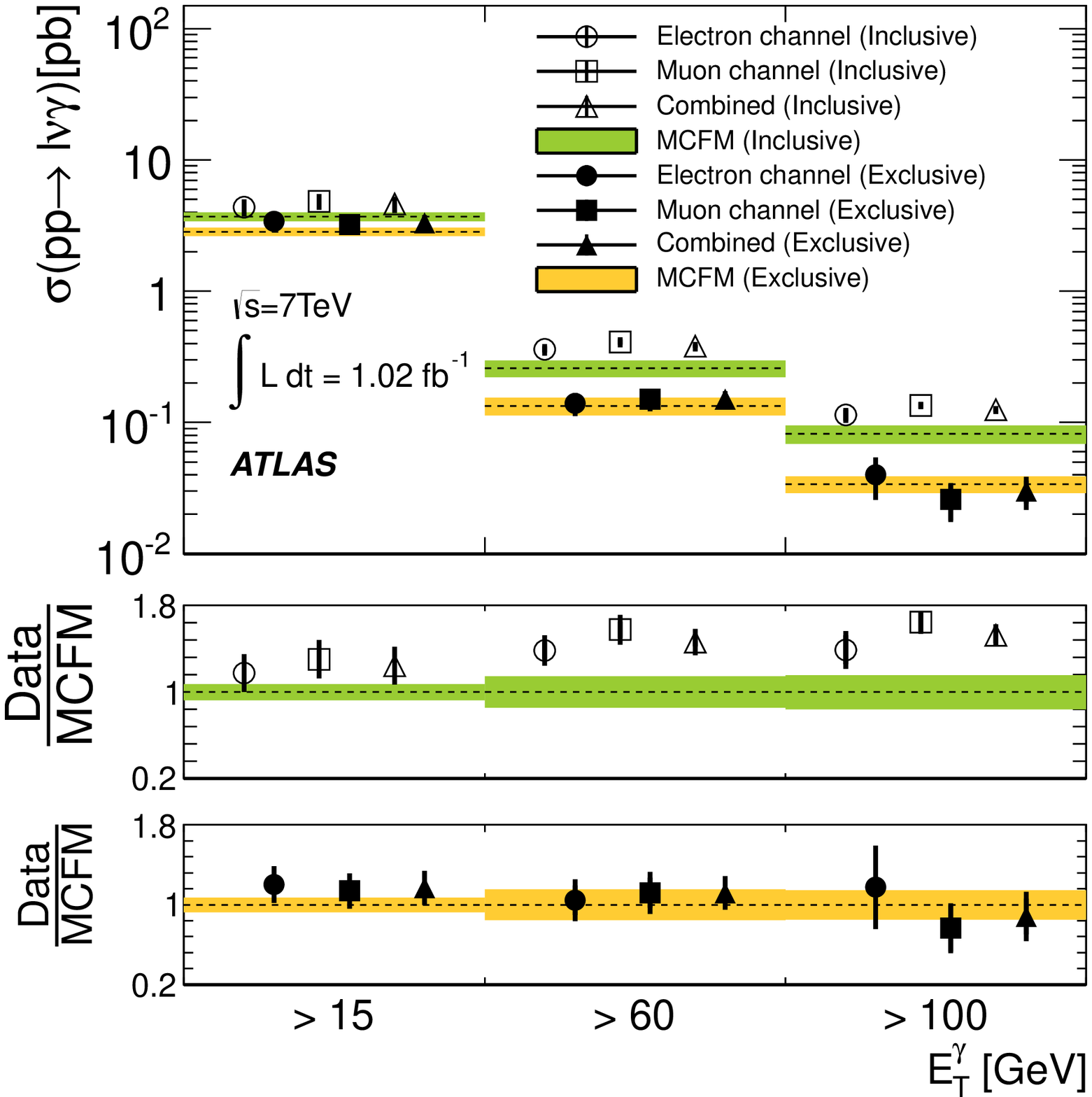,width=0.52\linewidth}%
    \psfig{figure=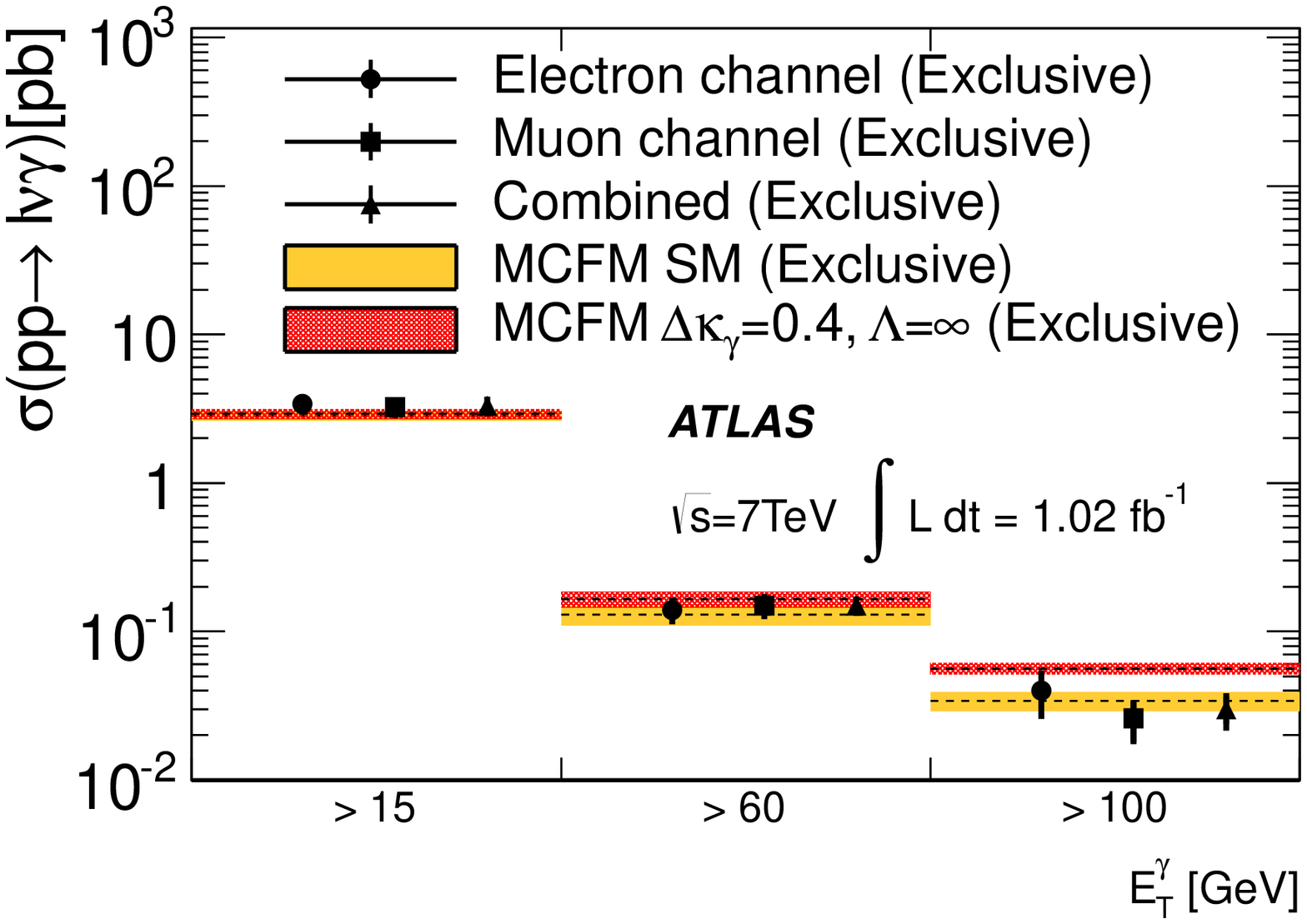,width=0.48\linewidth}
  \end{center}
  \caption{Left: Cross section measurement for the production of
    $W\gamma$ di-boson final states. Separate electron and muon and
    combined measurements are shown for three different thresholds of
    photon transverse momentum as well as for inclusive and exclusive
    selection. The measurements are compared to the MCFM prediction,
    where an overall good agreement is observed only for the exclusive
    selection~\protect\cite{Aad:2012mr}. Right: Anomalous triple gauge couplings predict an
    increase of cross section mostly at high photon transverse
    momentum~\protect\cite{Aad:2012mr_web}. \label{fig:wgamma}}
\end{figure}

The highest cross section of the di-boson processes is expected for
$W\gamma$ and $Z\gamma$ production. Here an additional well isolated
photon ($\Delta R(\ell,\gamma) > 0.7$) with high transverse momentum
(typically $E_T^\gamma>15\,$GeV) is selected together with a $W$ or $Z$
candidate. Cross sections have been measured with $\sim
10-15\%$ uncertainty~\cite{Aad:2012mr,Chatrchyan:2011rr}. One example
for cross sections measured in the $W\gamma$ channel is shown in
\Fig~\ref{fig:wgamma}. The agreement with NLO
prediction depends significantly on the
momentum cut applied to the photon as well as the presence of
additional jets in the event. For low photon momentum selection
$E_T^\gamma>15\,$GeV the agreement is reasonable irrespective of the
jet activity. But for high $E_T^\gamma>100\,$GeV the agreement is only
good for an ``exclusive selection'', where events with additional jets
are vetoed. Anomalous TGCs would enhance the cross section for high
$E_T^\gamma$.

\begin{figure}
  \begin{center}
    \psfig{figure=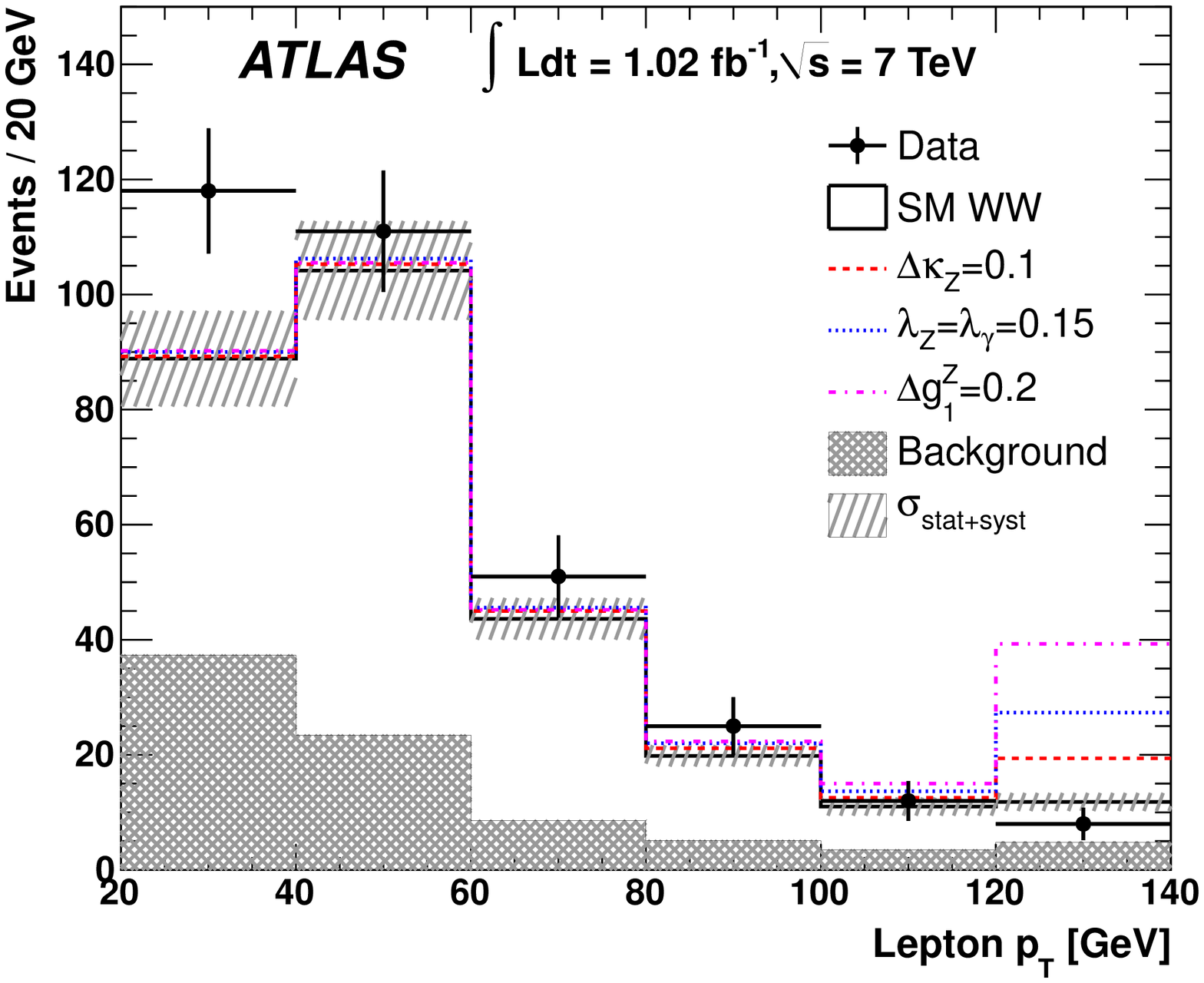,width=0.5\linewidth}
    \psfig{figure=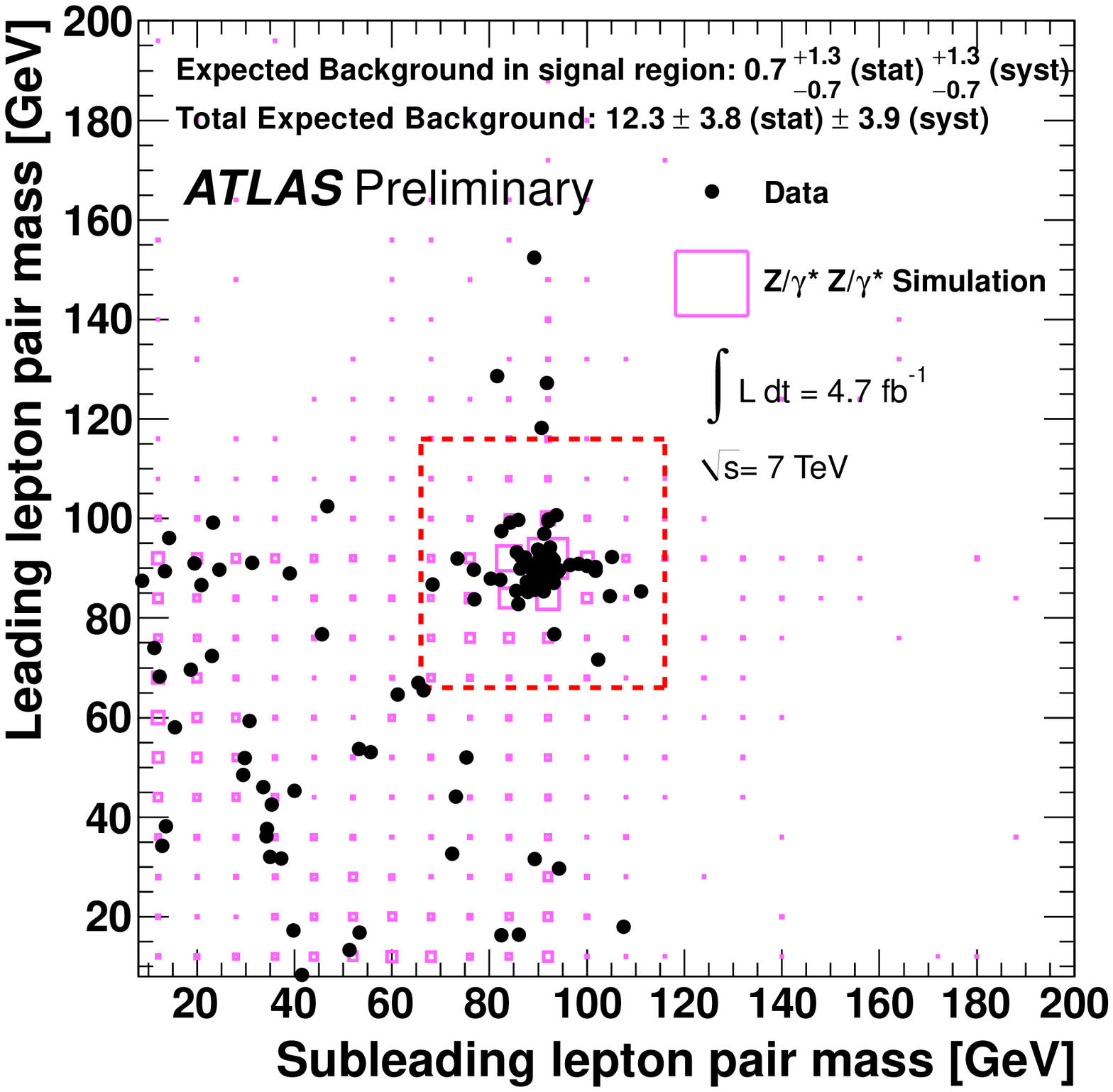,width=0.45\linewidth}
  \end{center}
  \caption{Left: Leading lepton transverse momentum in $WW$
    production, which is seen to agree well with the SM signal plus
    background expectation (solid lines). Non-standard TGCs would lead
    to a modification of the cross section mostly at high momentum as
    indicated by the coloured dashed and dotted lines~\protect\cite{Aad:2012ks}. Right:
    Candidates for doubly-resonant $ZZ$ production. A clear
    enhancement of candidates with very small background expectation
    can be seen in the signal box~\protect\cite{ZZ2011}.\label{fig:ww_zz}}
\end{figure}

The production of $WW$ final states is measured in the leptonic decay
channels with two charged leptons (electron or muon) and two
neutrinos. The background background from similar final states, i.e.
single $Z/\gamma^*$ Drell-Yan and top quark decays, is quite
significant. It can be controlled exploiting the missing transverse
momentum carried by the neutrinos, vetoing on di-lepton masses near
the $Z$ resonance and additional jets typical for $t\bar{t}$
production. Also here cross sections have been measured with
$\sim 10-15\%$ uncertainty~\cite{Aad:2012ks,WW2011,CMS-PAS-EWK-11-010}. 
The ATLAS and CMS results are both about $1\sigma$ above the NLO prediction.
Limits on aTGC are mostly sensitive to the leading lepton high $p_T$
tail, as can be seen in \Fig~\ref{fig:ww_zz}.

The final measurements to be discussed are for $WZ$ and $ZZ$
production. The cross section times branching to the purely leptonic
channels is very small and just a few $10$ events are observed per
$1\fbi$. The background is however very small with essentially no SM
processes producing events with 3 or 4 high $p_T$ isolated leptons. A
highly efficient lepton identification is mandatory to improve the
event yield, as the current accuracy is most limited by candidate
statistics reaching about $\sim 15\%$. The measurements are consistent
with the SM NLO
prediction~\cite{Aad:2011cx,Aad:2011xj,ZZ2011,CMS-PAS-EWK-11-010}.
Figure~\ref{fig:ww_zz} right shows the distribution of $ZZ$ candidates
as function of the leading and subleading di-lepton masses. A clear
enhancement in the $Z$ resonance region is observed, as expected for
$ZZ$ production. CMS has also considered $ZZ$ final states with
hadronic tau decays, where $1$ additional event is observed and
expected for $1\fbi$. ATLAS has measured $ZZ$ final states, where one
$Z$ decays invisibly to two neutrinos~\cite{ZZllnunu2011}. While this
has a higher branching ratio, the $Z$ bosons must have a significant
transverse momentum to be detectable as missing transverse momentum.
The analysis suffers from larger backgrounds similar to the $WW$
analysis and a strong missing transverse momentum cut and jet
veto are needed to control backgrounds. The measurement has currently $\sim 30\%$
total uncertainty and is consistent with the SM prediction and $ZZ \to
4\ell$ channel.

\section{Summary}

Thanks to the excellent performance of the LHC and the ATLAS and CMS
detectors, there is a wealth of high precision electroweak
measurements already 2 years after the data taking has started. The high
production rate of single $W$ and $Z$ bosons have enabled the
collaborations to do many detailed studies, among which are various
precise differential cross sections or production in association with
one or more light or heavy flavour jets. It has been shown, that
measurements using the first year of LHC data has an impact on our
knowledge of proton structure, e.g. the magnitude of the strange quark
density. Also the measurements of fundamental parameters of the
electroweak sector of the Standard Model are advancing, e.g. of the
weak mixing angle $\sin^2\theta_W$.

Finally, the di-boson measurements have quickly surpassed the
``observation phase''. The measurements are now testing the
non-Abelian gauge structure and limits on anomalous triple gauge
couplings are derived, which are of comparable quality to LEP and
Tevatron results.

Given the continued excellent performance of the LHC also in 2012,
many more interesting results can be expected in the near future.


\section*{References}
\begin{thebibliography}{99}
\bibitem{Aad:2008zzm}
  ATLAS Collaboration,
  JINST {\bf 3} (2008) S08003.
\bibitem{Chatrchyan:2008aa}
  CMS Collaboration,
  JINST {\bf 3} (2008) S08004.
\bibitem{Aad:2011dm}
  ATLAS Collaboration,
  Phys.\ Rev.\ D {\bf 85} (2012) 072004
  [arXiv:1109.5141 [hep-ex]].
\bibitem{CMS:2011aa}
  CMS Collaboration,
  JHEP {\bf 1110} (2011) 132
  [arXiv:1107.4789 [hep-ex]].
\bibitem{Aad:2011kt}
  ATLAS Collaboration,
  Phys.\ Rev.\ D {\bf 84} (2011) 112006
  [arXiv:1108.2016 [hep-ex]].
\bibitem{Ztau2011}
  ATLAS Collaboration,
  ATLAS-CONF-2012-006, https://cdsweb.cern.ch/record/1426991.
\bibitem{Chatrchyan:2011nv}
  CMS Collaboration,
  JHEP {\bf 1108} (2011) 117
  [arXiv:1104.1617 [hep-ex]].
\bibitem{Aad:2011fu}
  ATLAS Collaboration,
  Phys.\ Lett.\ B {\bf 706} (2012) 276
  [arXiv:1108.4101 [hep-ex]].
\bibitem{CMS-PAS-EWK-11-019}
  CMS Collaboration,
  CMS-PAS-EWK-11-019, https://cdsweb.cern.ch/record/1403095.
\bibitem{Aad:2012cu}
  ATLAS Collaboration,
  Submitted to Eur. Phys. J. C
  [arXiv:1204.6720 [hep-ex]].
\bibitem{Chatrchyan:2011jz}
  CMS Collaboration,
  JHEP {\bf 1104} (2011) 050
  [arXiv:1103.3470 [hep-ex]].
\bibitem{CMS-PAS-EWK-11-005}
  CMS Collaboration,
  CMS-PAS-EWK-11-005, https://cdsweb.cern.ch/record/1377410.
\bibitem{Chatrchyan:2011wt}
  CMS Collaboration,
  Phys.\ Rev.\ D {\bf 85} (2012) 032002
  [arXiv:1110.4973 [hep-ex]].
\bibitem{CMS-PAS-EWK-11-007}
  CMS Collaboration,
  CMS-PAS-EWK-11-007, https://cdsweb.cern.ch/record/1439026.
\bibitem{Chatrchyan:2012xt}
  CMS Collaboration,
  Submitted to Phys. Rev. Lett. [arXiv:1206.2598 [hep-ex]], CERN-PH-EP-2012-151.
\bibitem{Aad:2012sb}
  ATLAS Collaboration,
  Accepted by Phys. Rev. Lett.
  [arXiv:1203.4051 [hep-ex]].
\bibitem{Aaron:2009aa}
  H1 and ZEUS Collaboration,
  JHEP {\bf 1001} (2010) 109
  [arXiv:0911.0884 [hep-ex]].
\bibitem{CMS-PAS-EWK-11-013}
  CMS Collaboration,
  CMS-PAS-EWK-11-013, https://cdsweb.cern.ch/record/1369558.
\bibitem{Aad:2011kp}
  ATLAS Collaboration,
  Phys.\ Lett.\ B {\bf 707} (2012) 418
  [arXiv:1109.1470 [hep-ex]].
\bibitem{Aad:2011jn}
  ATLAS Collaboration,
  Phys.\ Lett.\ B {\bf 706} (2012) 295
  [arXiv:1109.1403 [hep-ex]].
\bibitem{CMS-PAS-SMP-12-003}
  CMS Collaboration,
  Submitted to JHEP [arXiv:1204.1643 [hep-ex]], CERN-PH-EP-2012-049.
\bibitem{Aad:2011qv}
  ATLAS Collaboration,
  Phys.\ Rev.\ D {\bf 85} (2012) 032009
  [arXiv:1111.2690 [hep-ex]].
\bibitem{Aad:2012en}
  ATLAS Collaboration,
  Phys.\ Rev.\ D {\bf 85} (2012) 092002
  [arXiv:1201.1276 [hep-ex]].
\bibitem{Aad:2011xn}
  ATLAS Collaboration,
  Phys.\ Lett.\ B {\bf 708} (2012) 221
  [arXiv:1108.4908 [hep-ex]].
\bibitem{Chatrchyan:2011ne}
  CMS Collaboration,
  JHEP {\bf 1201} (2012) 010
  [arXiv:1110.3226 [hep-ex]].
\bibitem{ATLAS:2012au}
  ATLAS Collaboration,
  Eur.\ Phys.\ J.\ C {\bf 72} (2012) 2001
  [arXiv:1203.2165 [hep-ex]].
\bibitem{Chatrchyan:2011ig}
  CMS Collaboration,
  Phys.\ Rev.\ Lett.\  {\bf 107} (2011) 021802
  [arXiv:1104.3829 [hep-ex]].
\bibitem{Chatrchyan:2011ya}
  CMS Collaboration,
  Phys.\ Rev.\ D {\bf 84} (2011) 112002
  [arXiv:1110.2682 [hep-ex]].
\bibitem{Aad:2012mr}
  ATLAS Collaboration,
  Submitted to Phys.\ Lett.\ B [arXiv:1205.2531 [hep-ex]].
\bibitem{Aad:2012mr_web}
  ATLAS Collaboration,
  https://atlas.web.cern.ch/Atlas/GROUPS/PHYSICS/PAPERS/STDM-2011-26.
\bibitem{Chatrchyan:2011rr}
  CMS Collaboration,
  Phys.\ Lett.\ B {\bf 701} (2011) 535
  [arXiv:1105.2758 [hep-ex]].
\bibitem{Aad:2012ks}
  ATLAS Collaboration,
  Phys.\ Lett.\ B {\bf 712} (2012) 289
  [arXiv:1203.6232 [hep-ex]].
\bibitem{WW2011}
  ATLAS Collaboration,
  ATLAS-CONF-2012-025, http://cdsweb.cern.ch/record/1430734.
\bibitem{CMS-PAS-EWK-11-010}
  CMS Collaboration,
  CMS-PAS-EWK-11-010, https://cdsweb.cern.ch/record/1370067.
\bibitem{Aad:2011cx}
  ATLAS Collaboration,
  Phys.\ Lett.\ B {\bf 709} (2012) 341
  [arXiv:1111.5570 [hep-ex]].
\bibitem{Aad:2011xj}
  ATLAS Collaboration,
  Phys.\ Rev.\ Lett.\  {\bf 108} (2012) 041804
  [arXiv:1110.5016 [hep-ex]].
\bibitem{ZZ2011}
  ATLAS Collaboration,
  ATLAS-CONF-2012-026, http://cdsweb.cern.ch/record/1430735.
\bibitem{ZZllnunu2011}
  ATLAS Collaboration,
  ATLAS-CONF-2012-027, http://cdsweb.cern.ch/record/1430736.

\end{thebibliography}
\end{document}